\documentclass[twocolumn,showpacs,preprint]{revtex4}

\usepackage{amsmath}

\usepackage{graphicx}

\begin{document}
\draft

\title{Sound beyond the speed of light: destructive interference, anomalous
dispersion and nonlocality of near field }
\author{Mark E. Perel'man *}
\affiliation{The Racah Institute of Physics, the Hebrew University,
91904 Jerusalem, Israel}

\begin{abstract}
Experimentally fixed sound pulse beyond of light speed in the region
of anomalous dispersion [W. M. Robertson, \textit{et al.} Appl.
Phys. Lett, 90, 014102 (2007)] can be explained, as well as the
similar superluminal phenomena, by "the nonlocality in the small" of
near electromagnetic field at transferring of relevanted
excitations.
\end{abstract}
\date{\today}
\maketitle


In last years numerous experiments (e.g., the reviews~\cite{GN}) had
been shown that the electromagnetic excitation or signal can be
passed in substance with a speed surpassed the light speed in
vacuum. The most contra-intuitive experiment~\cite{WM}, as if at the
special contradiction to the common sense, has shown, that the speed
faster-then-$c$ is possible even for a sound in the region of
anomalous dispersion in substances of normal density. (It is
necessary to emphasize, that the majority of experiences of so named
superluminal phenomena also were carried out in the regions of
anomalous dispersion.)

Basic distinction between~\cite{WM} and experiments in optical area
or in electronic devices consists in the following: if the excess of
speed in them was not so great and it yet was possible to think of
rearrangement of waves' envelope and so on, the speed excess in
acoustics is of the six orders (!) and consequently it become
evident that a more radical approach is required. Hence the adequate
approach that can simultaneously explain both optical and acoustic
experiments must be developed and verified, especially since it can
be assumed that in the processes of superfast transferring of sound
the electromagnetic phenomena can take part or even play a basic
role.

It can be assumed that common complexities at consideration of these
problems are connected with the examinations of solutions of the
wave equations in the beforehand supposed form
$f(\mathbf{v}t-\mathbf{r})$, i.e. as the traveling waves only, that,
certainly, demands performance of the condition $v\leq c$. In a near
field, however, can exist also faster decreasing nonlocal solutions
(e.g.~\cite{MEP}). Therefore other approaches, without a priori
introduced restrictions in the analysis of wave equations and/or of
initial conditions, must be examined.

In the cited experiments of the Robertson group~\cite{WM} anomalous
dispersion is created by destructive interference when the path
length $\Delta L$ between the long and short arms of sound circuit
differs by an odd number of one-half wavelengths. (In early
researches in this area the article~\cite{NSS} must be noted, but
with more complex substance that complicates precise conclusions.)

Let's consider some details of the experiment. Oscillations' velocity of
particles in sound field $v=p/\rho s$, where $p$ is the acoustic pressure, $%
\rho $ is the density of substance, $s$ is the speed of sound. With stopping
of atoms participating in sound process by destructive interference, each of
them would emit the electromagnetic momentum $mv=\hbar k$ in the direction
of sound pulse, that at the real photon emission may correspond to the
energy $E=\hbar ck=\hbar \omega $.

But the energy, emitted at considered stopping, $E_1=\hbar^2k^2/2m
$, does not correspond to the momentum of real photon. Therefore the
forward transferring of this momentum can go only as the tunneling.
The deficit of energy

\begin{equation}
\Delta E=\hbar ck-(\hbar k)^{2}/2m=\hbar \omega \left( 1-\hbar \omega
/2mc^{2}\right)   \label{(1)}
\end{equation}

is equal to the height of potential barrier that must be overcoming by
virtual photon till its possible absorption by subsequent atom on a distance
$\Delta L$. This distance coincides in accordance with the uncertainty
principle with $\hbar c/\Delta E$, i.e. with corresponding wave length.

How this process can be represented in the frame of QED? The
propagator of near field can be defined via the decomposition of
general propagator of quantum electrodynamics onto far, intermediate
and near fields~\cite{MEP}:

\begin{equation}
D_{NF}(t,\mathbf{r})=\frac{1}{4\pi r^{3}}\left[ r\text{sgn}(t)\theta
(t^{2}-r^{2})+t\theta (r^{2}-t^{2})\right]
\end{equation}

(here and below $\hbar =c=1$, projectors of polarization are omitted). The
propagator contains local and nonlocal parts, and its complete Fourier
transform shows that just the nonlocal part corresponds to the region of
anomalous dispersion:

\begin{equation}
D_{NF}(\omega ,\mathbf{k})=\frac{1}{8\pi ^{2}i\omega ^{2}k}\left[ \left\vert
\omega \right\vert \theta (\omega ^{2}-k^{2})+k\theta (k^{2}-\omega ^{2})%
\right]
\end{equation}

Let's consider this part in the ($\omega $, $\mathbf{r}$)-representation:

\begin{equation}
D_{NF}(\omega ,\mathbf{r})=\frac{\pi }{i\omega ^{2}}\delta (\mathbf{r})-%
\frac{1}{2\pi i\omega ^{2}r^{3}}\left[ \sin \omega r-\omega r\cos \omega r%
\right]
\end{equation}

The first term in (4) is related to the contact interaction and in our
analysis must be omitted. The second term is reducing at $\omega
r\rightarrow 0$ to

\begin{equation}
D_{NF}(\omega ,\mathbf{r})\propto \frac{\omega }{6\pi }
\end{equation}

and remains approximately constant till $\omega r\sim \pi /2$
whereupon appreciably and sufficiently promptly decreases. It means
a nonlocal transmission of excitations on definite distances only.

The nonlocality of transfer of excitation, shown in such qualitative
description, can be investigated by the deeper and the most general
methods of dispersion relations based on the general requirements of
causality, spectrality and so on (e.g.~\cite{MEP1}). It does not
require dynamical considerations and leads to the restriction by
only the kinematical arguments.

So, in the article~\cite{MEP2} (more detailed in~\cite{MEP1}) we had
investigated this problem in the frame of general covariant
dispersion relations. But for all that, what type of excitation
(electromagnetic, sound or gravitational) can be nonlocally
transferred and its possible speed were not predetermined: the
proved theorem is of general character. It had been shown that in
the region of anomalous dispersion an instantaneous transferring
(instant jumps) of excitations is possible only within scopes of
near field on the distances $\Delta L$ determined by wavelength
corresponding to the energy deficiency $\Delta E$ relative to the
nearest stable (resonance) state:

\begin{equation}
\Delta L=\frac{\pi c\hbar }{\Delta E}(2n-1);\qquad \qquad n=1,2,3,\ldots \ .
\end{equation}

This expression completely corresponds to the results of
experiments~\cite{WM} and allows to state that in considered
experiments really was observed the instantaneous transferring of
sound excitation onto strongly definite distances by near
electromagnetic field.

Such conclusion seems, at first sight, excessively courageous,
especially with the basic attracting of arguments, initially
developed for electromagnetic theory, to acoustics. However, such
reasons can be considered. Firstly, in the basis of all sound
phenomena are electromagnetic interactions of particles of
substance; moreover we can think that the interactions between
constituents of condensed media can take place in the anomalous
dispersion fields and probably can be instantaneous~\cite{MEP3}.
Second, the results of some well known "superluminal" experiments
[8] can be explained by the existence of conditions of anomalous
dispersion aroused by destructive interference of rays passing
through multi-layer interference filters.

So, the considered phenomenon represents quantum tunneling or "the
nonlocality in the small" and is connected to particles (waves) with excess
of momentum relative to energy. Notice that even at $n=1$ the value of (6)
is twice bigger the uncertainties value and therefore is measurable. If
measurements of duration of a signal passage are made on distances
sufficiently bigger $\Delta L$, i.e. they include both the distance of
instant jump of excitation and distances with the normal speed of
propagation, then, certainly, any values of signal speeds, including
superluminal ones, can be received.

Notice that the considered emission of suddenly stopped atoms can be
examined as the atomic bremsstralung (cf.~\cite{MYA}). It is known
that just in processes of bremsstralung was discovered the
phenomenon of gradual formation (dressing) of photons, i.e. the
necessity of certain time (path) for their completion, till which
they remain virtual ones~\cite{MLT}. This phenomenon is actively
investigating in the high energy physics (e.g.~\cite{VNB}), but must
be appreciable at very low energies also, at description of van der
Waals interactions also~\cite{MEP3}. Note that this result can be
especially interesting for astrophysics and cosmology
(e.g.~\cite{GFR} and references therein), but they are far from the
purpose of this letter.

In conclusion it is possible to maintain that the analysis of
experimental researches shows the nonlocality of interactions in
near fields under certain conditions and on strictly definite
distances only. Such conclusion does not contradict the basis of
relativity considering far fields and examined only in them. We
stress that our description corresponds to Wigner's formulation of
common causality: \textit{The scattered wave cannot escape a
scatterer before the initial wave reaches it}~\cite{WIG}.

*). E-mail: mark-perelman@mail.ru

\end{document}